\documentclass[a4paper,11pt]{spieman}  
\usepackage{amsmath,amsfonts,amssymb}
\usepackage{graphicx}
\usepackage{setspace}
\usepackage{tocloft}
\usepackage{booktabs}
\usepackage{multirow}
\usepackage{physics}
\usepackage{textcomp}
\DeclareMathOperator{\sinc}{sinc}

\newcommand{\red}{\textcolor{black}}

\title{Interferometric Beam Combination with a Triangular Tricoupler Photonic Chip}

\author[a,*]{Jonah T. Hansen}
\author[a]{Michael J. Ireland}
\author[b]{Andrew Ross-Adams}
\author[b]{Simon Gross}
\author[c]{Tiphaine Lagadec}
\author[c]{Tony Travouillon}
\author[c]{Joice Mathew}
\affil[a]{Research School of Astronomy and Astrophysics, College of Science, Australian National University, Canberra, Australia}
\affil[b]{MQ Photonics Research Centre, Department of Physics and Astronomy, Macquarie University, Sydney, Australia}
\affil[c]{Advanced Instrumentation Technology Centre, Research School of Astronomy and Astrophysics, College of Science, Australian National University, Canberra, Australia}

\cftpagenumbersoff{figure}
\cftpagenumbersoff{table} 
\begin{document} 
\maketitle

\begin{abstract}
Beam combiners are important components of an optical/infrared astrophysical interferometer, with many variants as to how to optimally combine two or more beams of light to fringe-track and obtain the complex fringe visibility. One such method is the use of an integrated optics chip that can instantaneously provide the measurement of the visibility without temporal or spatial modulation of the optical path. Current asymmetric planar designs are complex, resulting in a throughput penalty, and so here we present developments into a three dimensional triangular tricoupler that can provide the required interferometric information with a simple design and only three outputs. Such a beam combiner is planned to be integrated into the upcoming \textit{Pyxis} interferometer, where it can serve as a high-throughput beam combiner with a low size footprint. Results into the characterisation of such a coupler are presented, highlighting a throughput of \red{85$\pm7$\%} and a flux splitting ratio between 33:33:33 and 52:31:17 over a 20\% bandpass. We also show the response of the chip to changes in optical path, obtaining an instantaneous complex visibility and group delay estimate at each input delay. 
\end{abstract}


{\noindent \footnotesize\textbf{*}Email:  \linkable{jonah.hansen@anu.edu.au} }

\begin{spacing}{1}   

\section{Introduction and Motivation}
\label{sect:intro}  

The beam combiner in any optical interferometer is a critical component, performing the duties of combining the electric fields and producing the fringes that encode astrophysical data. However, how to optimally do this in a photon-starved environment as is the case for most interferometers is a complex and application-specific question. 

When it comes to combining the light from multiple telescopes, there are a number of ways to achieve this  \cite{Buscher2015Practical,minardi_beam_2016} . One can spatially encode the fringes as in Young's double slit experiment; that is to directly combine the light on the detector. Beam combiners using this encoding include \textit{PAVO} \cite{2008Ireland} \red{and \textit{MIRC-X} \cite{MIRCX} on the \textit{CHARA} array,} and \textit{MATISSE} \cite{lopez_overview_2014} and the now decommissioned \textit{AMBER} \cite{petrov_amber_2007} on the Very Large Telescope Interferometer (\textit{VLTI}). 

Another way to ``encode'' the fringe is known as temporal encoding. Here, the light from the two arms of the interferometer are co-axially combined at a beamsplitter, resulting in a beam with uniform intensity. This beam can then be detected with a single pixel detector. To reconstruct the fringe, if one arm of the interferometer is pistoned with time then the phase difference of the two arms of the interferometer will change and so the intensity on the detector will vary sinusoidally with time. This has the advantage of having a higher signal to noise in a readout noise limited regime due to using one pixel, but has the disadvantage that, in addition to mechanical complexity, multiple measurements have to be made within a coherence time due to the modulation of the optical path. The \textit{COAST} \cite{haniff_coast_2004}~, \textit{SUSI} \cite{2008SPIE.7013E..04T} and \textit{NPOI} \cite{armstrong_navy_1998} interferometers have used such a scheme to encode the fringes.

Recently, advances in photonics have led to guided light combiners using single mode waveguides with numerous advantages. First, if single mode fibers are used for propagation, then the fibers spatially filter out modes caused by the turbulent atmosphere, leaving the fundamental mode to be combined coherently \cite{coude_du_foresto_deriving_1997}~. In this way, turbulence does not affect the ability to interfere the light, but instead affects the amount of light that can be injected into the fibers.  One of the first beam combiners to use optical fibers was \textit{FLUOR} \cite{coude_du_foresto_fluor_1998}~, which used a 2x2 optical fiber coupler instead of a beam splitter. However, the two outputs leave an ambiguity as to the sign of the phase, and so still required the use of temporal delay modulation.

Integrated optics (IO) allow optical waveguides to be etched into a small piece of glass or crystal, essentially the optical equivalent of integrated circuits \cite{Buscher2015Practical}~. Complex guides can be built onto a single chip, which has led to great leaps in beam combination. For example, the \textit{GRAVITY} \cite{abuter_first_2017} and \textit{PIONEER} \cite{bouquin_pionier_2011} beam combiners for the \textit{VLTI} use a complex IO chip (known as an ABCD combiner) to produce four outputs for each pair of telescopes. As the \textit{VLTI} utilises four telescopes with 6 baselines between them, the chip has 24 outputs in total. The outputs (denoted $A$, $B$, $C$ and $D$) measure the light with notional phase offsets of 0, 90, 180 and 270~degrees respectively, which can then reconstruct the complex coherence without the use of temporal fringe scanning. \red{One such simple ABCD estimator (used by the Mark III interferometer \cite{1988A&A...193..357S} and the Palomar Testbed Interferometer \cite{1999ApJ...510..505C}) is:}
\begin{align}
    \gamma = \frac{A-C + i(B-D)}{A+B+C+D}.
\end{align}

This is an advantage over other combination schemes since a lack of modulation means that the entire integration time can be spent on measurement. Furthermore, these IO chips can be made to be very small and so work very well under tight volume constraints. The downside to this method is the complexity in the chip, which reduces the throughput of the beam combiner and also can be quite expensive at wavelengths outside of an optical communications band. The desired phase shift in this type of IO chip can also be difficult to achieve \cite{2009A&A...498..601B}~. 

Here, we present a waveguide architecture that can obtain the complex visibility using a far simpler photonic chip and with \red{one fewer output} when compared to the ABCD combiner. This architecture, known as the triangular tricoupler, has three inputs and three outputs with a phase offsets of $\pm\frac{2\pi}{3}$. This results in a higher throughput per pixel on the detector, as the light is split into fewer pixels, while still providing complete information (due to the phase offsets) on the complex coherence.  Such a chip has seldom been used in a interferometer before, although there have been explorations into planar tricouplers for stellar interferometry \cite{labeye_integrated_2004,lacour_new_2014}~, and such devices are already implemented in quantum optics applications \cite{chung_broadband_2012}~.

\red{Our motivation for developing this chip is for use as a visible wavelength beam combiner for the \textit{Pyxis} project, a ground prototype for a formation flying interferometer \cite{hansen_ireland_2020,pyxis}. This project is part of a series of testbeds designed to push forward with space interferometry, with the aim of eventually producing a large scale mission that is able to detect Earth-like exoplanets around Solar type stars (eg. Large Interferometer For Exoplanets \cite{LIFE,LIFE1}). \textit{Pyxis}, being designed to be fit within a CubeSat footprint, has tight space restrictions and benefits from a lack of moving parts. As such, a photonic chip with fewer outputs and therefore increased signal is an ideal choice. In addition to being a testbed, Pyxis will have the capability for unique astrophysical measurements, including precise differential interferometric polarimetric measurements of mass losing giant stars, constraining dust formation, which requires 2-20m baselines at a 700\,nm wavelength \cite{2005MNRAS.361..337I,2012Natur.484..220N}.}

This paper is structured as follows: section \ref{sect:theory} will briefly describe the theory behind the architecture, section \ref{sect:manufacturing} will describe the manufacturing process in creating the IO chip and section \ref{sect:results} will detail the characterisation and performance of the chip in estimating visibilities and group delay for fringe tracking.

\section{Theoretical Development}
\label{sect:theory}

A tricoupler is fundamentally a device that makes a unitary transform between three electric fields to another three electric fields. We are considering primarily a situation where it is used as a beam combiner with two input beams (with no light injected into the third input) \cite{2010ApOpt..49.6675H} and not as a beam splitter \cite{lacour_new_2014}~. There are only certain design transformations that are possible. We can impose several restrictions for an intuitively ideal combiner:

\begin{enumerate}
    \item Left-right symmetry.
    \item Unitary transform, i.e. a lossless coupler.
    \item Balanced intensity on all three outputs when injected into one input
\end{enumerate}

We'll demonstrate the development of the coupling transfer matrix using a directional coupler before examining the three way coupler.

\subsection{Two Way Coupler}

Using the scalar approximation of the electric field, the coupling equation of waveguides in a photonic coupler, as stated in Snyder and Love's ``Optical Waveguide Theory'' \cite{snyder2012optical} \red{(or alternatively the review by Huang 1994 \cite{1994JOSAA..11..963H})}, is given by:
\begin{equation}
    \dv{\vb{b}}{z} = i\vb{A}\vb{b}
\end{equation}
where
\begin{align}
   \vb{A} = \begin{bmatrix}
    \beta_0+C_{11} & C_{12}\\
    C_{21} & \beta_0 + C_{22}
    \end{bmatrix},
\end{align}
$\beta_0 = \kappa n_0$ is the propagation constant for both fibers, $\kappa$ is the angular wavenumber, $C_{i,j}$ are the coupling coefficients and $n_0$ is the effective refractive index. Making the substitution $c_{ij} = \frac{C_{ij}}{\beta_0}$ we then have:
\begin{align}
   \vb{A} = \beta_0\vb{A'} = \beta_0 \begin{bmatrix}
    1+c_{11} & c_{12}\\
    c_{21} & 1 + c_{22}
    \end{bmatrix} \approx \beta_0 \begin{bmatrix}
    1 & \delta\\
    \delta & 1
    \end{bmatrix}
\end{align}
where we have ignored the diagonal $c_{11}$ and $c_{22}$ terms due to $c_{ii} << 1$ and set $\delta = c_{12} = c_{21}$ due to symmetry.

We assume a solution where $z$ is the coordinate along the propagation axis, 
\begin{align}
    \vb{b}_j\red{(z)} = \vb{v}_je^{i\beta_jz},
\end{align}
which produces an eigenvector equation with effective propagation constants:
\begin{align}
    \vb{A}'\red{\vb{v}_j} = \frac{\beta_j}{\beta_0}\red{\vb{v}_j} && \beta_j = \beta_0\{1+\delta,1-\delta\}
\end{align}
\red{and eigenvectors
\begin{align}
    \vb{v}_j = \left\{\frac{1}{\sqrt{2}}[1,1],\frac{1}{\sqrt{2}}[1,-1]\right\}
\end{align}}
The difference in the effective propagation constants is given by:
\begin{align}
    \Delta \beta_{\text{eff}} &= \kappa \Delta n_{\text{eff}} = \beta_0(1 + \delta - (1 - \delta)) = 2\kappa n_0  \delta\\
    \Rightarrow \delta &= \frac{\Delta n_{\text{eff}}}{2n_0}
\end{align}
Hence the coupling coefficients \red{are simply related to the difference in effective indices of the two different modes. One can measure this difference using a photonic simulator such as \textit{RSoft} \cite{RSoft}} which in turn can produce the ideal length of the coupler to equalise output flux.

\subsection{Ideal Tricoupler}

\begin{figure}
    \centering
    \includegraphics[width=0.7\linewidth]{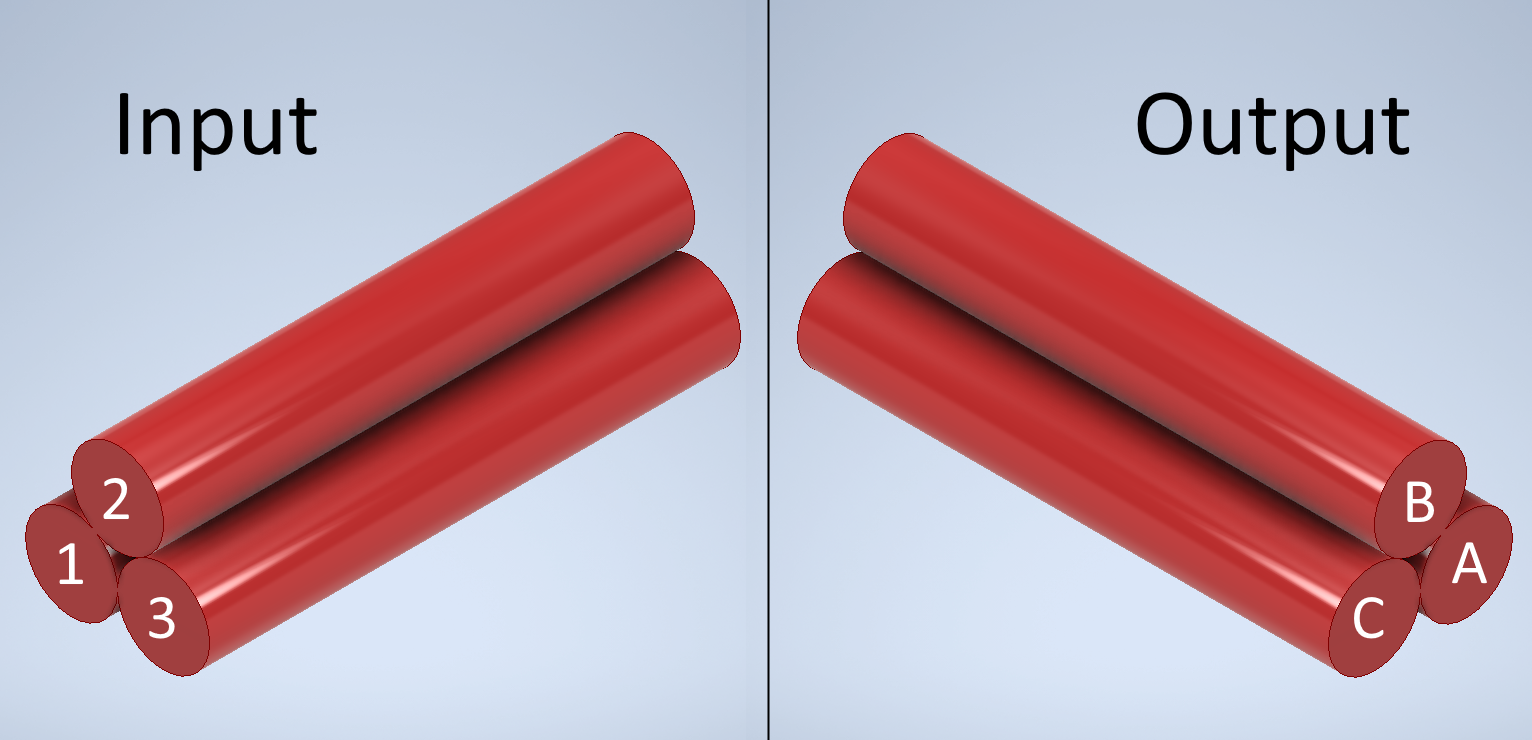}
    \caption{Simple isometric \red{view} of the triangle tricoupler, with labelled inputs \red{and outputs.} }
    \label{fig:tricoupler}
\end{figure}

We now turn to the tricoupler. An alternative treatment of the mathematics behind a 3x3 symmetrical coupler, \red{in terms of an application to optical communications}, can be found in Xie et. al (2012) \cite{Xie12}~. If we consider a naive planar coupler, the intrinsic three-way asymmetry requires an asymmetric layout (such as a larger central guide) in order to produce even coupling. This asymmetry can be resolved by instead considering a three dimensional equilateral triangle coupler, sketched in Figure \ref{fig:tricoupler}. This layout has rotational asymmetry, and as such can have all three guides being the same. The coupling matrix (calculated similarly to the two-way coupler above) of this layout is
\begin{align}
   \vb{A} = \beta_0\vb{A'} = \beta_0 \begin{bmatrix}
    1 & \delta & \delta\\
    \delta & 1 & \delta \\
    \delta & \delta & 1
    \end{bmatrix}
\end{align}

The eigenvalues of this system are:
\begin{align}
    \beta_j = \beta_0\left\{1-\delta,1+2\delta,1-\delta\right\} 
\end{align}
with eigenvectors:
\begin{align}
    \vb{v}_j = \left\{\frac{1}{\sqrt{2}}[-1,1,0],\frac{1}{\sqrt{3}}[1,1,1],\frac{1}{\sqrt{6}}[-1,-1,2]\right\}
\end{align}
giving the solution of
\begin{equation}
    \vb{b}(z) = \sum_{j=1}^3 a_j\vb{v}_je^{i\beta_jz}.
\end{equation}

We can now identify the length of the coupler, $z_\text{len}$, such that if light is injected into one fiber, the output light is distributed evenly between all three. \red{On injection, we have $\vb{b}(0) = [1,0,0]$}. Solving this through row reduction gives coefficients
\begin{align}
    a_j = \left\{-\frac{1}{\sqrt{2}},\frac{1}{\sqrt{3}},-\frac{1}{\sqrt{6}}\right\},
\end{align}
and so by expanding $\vb{b}(z_\text{len})$ we obtain:
\begin{align}
    \vb{b}(z_\text{len}) &= \frac{1}{6}\left([4,-2,-2]e^{i\beta_0(1-\delta)z_\text{len}} + [2,2,2]e^{i\beta_0(1+2\delta)z_\text{len}}\right).
\end{align}

\red{On the output, for all outputs to be of equal intensity, we require that $|\vb{b}(z_\text{len})| = \frac{1}{\sqrt{3}}[1,1,1]$. If we let $z_\text{len} = \alpha\frac{2\pi}{\beta_0\delta}$, where $\alpha$ is some fraction, we can clearly see that} we require $\alpha$ such that
\begin{align}
    |2e^{-{2i\pi\alpha}} + e^{4i{\pi\alpha}}| = |-e^{-{2i\pi\alpha}} + e^{4i{\pi\alpha}}| = \sqrt{3}
\end{align}
\red{This requirement is satisfied for a minimum $\alpha$ value of $\frac{1}{9}$. Recall as well that the coupling coefficients are related to the difference in effective indices of the modes. Due to the symmetry of the coupler, there will only be one $\Delta n_{\text{eff}}$ for the set of guides, and is related to the coupling coefficients by:
\begin{align}
    \Delta n_{\text{eff}} &= n_0(1 + 2\delta - (1 - \delta)) = 3n_0\delta\\
    \delta &= \frac{\Delta n_{\text{eff}}}{3n_0}
\end{align}}
The length required is hence given by:
\begin{align}
    \label{eq:theoretical_len}
    z_\text{len} &= \frac{2\pi}{9\beta_0\delta} = \red{\frac{\lambda}{3\Delta n_{\text{eff}}}.}
\end{align}
That is, the ideal length of the coupler is determined only by the wavelength and the difference in effective indices of each mode, \red{which as before in principle can be calculated with a photonic simulator}. 

By defining
\begin{align}
\label{eq:coupling_matrix}
    \vb{M} = \red{\frac{1}{\sqrt{3}}}\begin{bmatrix}
    1 & e^{i\frac{2\pi}{3}}  & e^{i\frac{2\pi}{3}} \\
    e^{i\frac{2\pi}{3}}& 1 & e^{i\frac{2\pi}{3}}\\
    e^{i\frac{2\pi}{3}} & e^{i\frac{2\pi}{3}} & 1
    \end{bmatrix},
\end{align}
the final coupling equation can thus be given by (ignoring common phase terms):
\begin{align}
    \vb{b}(z_\text{len}) = \red{\vb{M}\vb{b}(0)}.
\end{align}

\red{From now on, we use the notation $\vb{I} = |\vb{b}(z_\text{len})|^2$ to denote the output intensity at the end of the coupler.}

\subsection{\red{Ideal interferometric} output}

\begin{figure}
    \centering
    \includegraphics[width=0.7\linewidth]{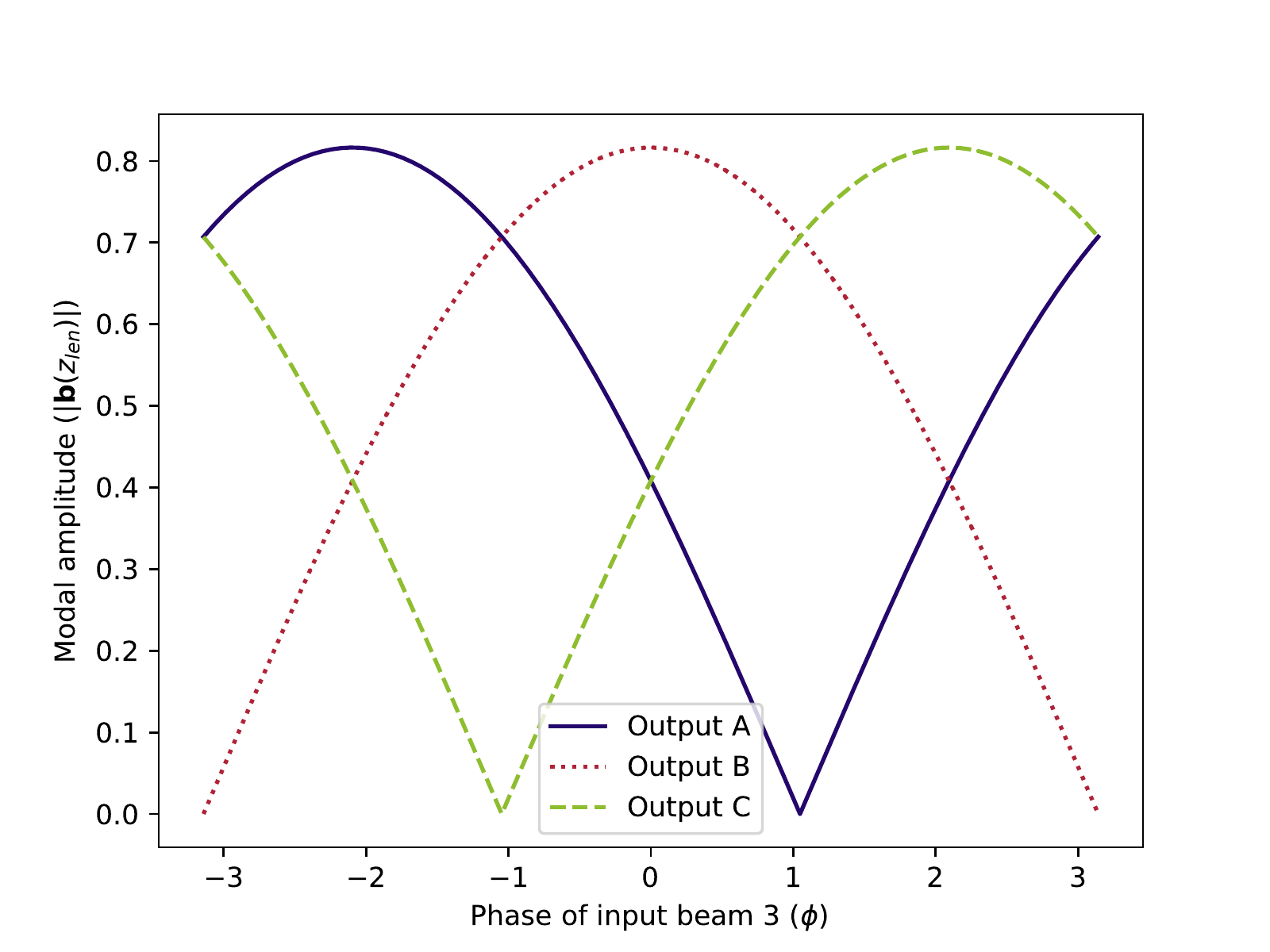}
    \caption{Output modal amplitude as a function of input relative phase difference for the triangular tricoupler. Note that each output is phase shifted by $\pm\frac{2\pi}{3}$.}
    \label{fig:phi_plot}
\end{figure}

In Figure \ref{fig:phi_plot}, we plot what happens if we have an input of $\vb{b}(0) = \frac{1}{\sqrt{2}}[1,0,e^{i\phi}]$ and then see how $|\vb{b}(z_\text{len})|$ changes as $\phi$ is varied. From this plot, we can see the tricoupler introduces a $\pm2\pi/3$ phase shift for two of the outputs.

Now, \red{the fringe intensity on a detector is simply a linear combination of the input fields, and so we can model the ideal intensity (in the absence of instrumental noise) in an analogous way as Buscher (2015) \cite{Buscher2015MakingFringes}. As a function of linear delay $x$, related to the phase by $\phi = \frac{2\pi x}{\lambda}$, the intensity is:}
\begin{equation}
\red{I}(x,\lambda) = F_0[1 + |\gamma|\cos(\frac{2\pi x}{\lambda} - \Phi)]
\end{equation}
where $F_0$ is the total flux from the object, $|\gamma| = V$ is the complex coherence modulus (also known as the visibility), $\lambda$ is the wavelength and $\Phi$ is the static phase delay from the astrophysical source corrupted by the atmosphere ($\Phi = \phi_{obj} + \phi_{atm}$). 

The intensities of the three outputs \red{($\vb{I} = [A,B,C]$)} are then given by
\begin{align}
A(x,\lambda) &= F_0\left[1 + V\cos(\frac{2\pi x}{\lambda} - \Phi)\right]\\
B(x,\lambda) &=F_0\left[1 + V\cos(\frac{2\pi x}{\lambda} - \Phi + \frac{2\pi}{3})\right]\\
C(x,\lambda) &= F_0\left[1 + V\cos(\frac{2\pi x}{\lambda} - \Phi + \frac{4\pi}{3})\right],
\end{align}
from which we can then derive the full complex coherence:
\begin{align}
    \gamma_i(x,\bar{\lambda}_i) &= \frac{3A_i+\sqrt{3}i(C_i-B_i)}{A_i+B_i+C_i} - 1
\end{align}

\subsection{The ``Pixel To Visibility Matrix'' (PV2M)}

Realistically, the tricoupler will not be ideal and so we must determine the coupling matrix experimentally. We can do this by injecting light into each single input $i$ and measuring the intensity for each output $j$ \red{(that is, $\mathbf{I}_{i,j}$ is the intensity of output $j$ when only input $i$ is illuminated)}. The normalised fractional intensity gives the squared modulus of the coupling matrix elements $\red{\mathbf{M}}_{i,j}$:
\begin{align}
    |\red{\mathbf{M}}_{i,j}|^2 = \frac{\red{\mathbf{I}}_{i,j}}{\sum_j\red{\mathbf{I}}_{i,j}}
\end{align}

To find the elements, we used a parameterisaton equivalent to the Kobayashi–Maskawa matrix in particle physics \cite{Kobayashi73}~. This parameterisation assumes a unitary (lossless) coupler, and is fully defined by four parameters: three ``mixing'' angles ($\theta_1,\theta_2,\theta_3$) and one phase angle $\delta$. The matrix is shown below. Note that $\cos{\theta_i}$ and $\sin{\theta_i}$ are denoted $c_i$ and $s_i$ respectively, \red{and this is a generalisation of equation \ref{eq:coupling_matrix}.}
\begin{align}
    \mathbf{M} = \begin{bmatrix}
    c_1 & -s_1c_1 & -s_1s_3\\
    s_1c_1 & c_1c_2c_3 - s_2s_3e^{i\delta} & c_1c_2s_3 + s_2c_3e^{i\delta}\\
    s_1s_2 & c_1s_2c_3 + c_2s_3e^{i\delta} & c_1s_2s_3 - c_2c_3e^{i\delta}
    \end{bmatrix}
\end{align}
Without loss of generality, we can restrict the mixing angles to be $0 \leq \theta \leq \frac{\pi}{2} $. The phase angle can be found through a least squares fit of the corresponding intensities. Note that there is no distinction between $\delta$ and $-\delta$, and so a slight \red{one-time callibration} phase modulation on the input guides \red{(on the order of $\pi/2$ radians)} is required to break this degeneracy.

Once the coupling matrix is defined, we can then define the output monochromatic intensity as a function of input phase:
\begin{align}
    \mathbf{\red{I}}(\phi) &= F_0 \mathbf{g}(\phi) = F_0\left|\mathbf{M}\cdot\frac{1}{\sqrt{2}}\begin{bmatrix}
    1\\
    0\\
    e^{i\phi}
    \end{bmatrix}\right|^2
\end{align}
where $\mathbf{g}(\phi)$ is the normalised intensity as a function of phase.

To find the complex coherence, we employ a similar method to previous photonic beam combiners \cite{2007NewAR..51..682T,2008SPIE.7013E..16L,2019A&A...624A..99L} in developing the ``visibility to pixel matrix'' (V2PM). The intensities at each of the three outputs can be written as a function of the real and imaginary part of the coherence ($\gamma$) through this matrix:
\begin{align}
    \red{\mathbf{I}} = \mathbf{V2PM}\cdot\begin{bmatrix}
    \mathfrak{R}(\gamma)F_0\\
    \mathfrak{I}(\gamma)F_0\\
    F_0
    \end{bmatrix}
\end{align}
where
\begin{align}
    \mathbf{V2PM} = \begin{bmatrix}
    \mathbf{r} & \mathbf{i} & \mathbf{f}
    \end{bmatrix}
\end{align}

These three vectors are derived from estimating the real and imaginary component of the visibilities, as well as the total flux. One easy way to estimate these vectors from a simulation is through the normalised intensity at a variety of phases:
\begin{align}
    \mathbf{r} &= \frac{1}{2}\left( \mathbf{g}(0) - \mathbf{g}(\pi)\right)\\ 
    \mathbf{i}&= \frac{1}{2}\left(\mathbf{g}\left(\frac{\pi}{2}\right) - \mathbf{g}\left(\frac{3\pi}{2}\right)\right)\\ 
    \mathbf{f}&= \frac{1}{4}\left(\mathbf{g}(0) + \mathbf{g}\left(\frac{\pi}{2}\right)+ \mathbf{g}\left(\pi\right)  + \mathbf{g}\left(\frac{3\pi}{2}\right)\right)
\end{align}

We can then invert this matrix, now named the ``pixel to visibility matrix'' (P2VM), to determine the complex coherence from a set of intensity measurements:
\begin{align}
    \begin{bmatrix}
    \mathfrak{R}(\gamma)F_0\\
    \mathfrak{I}(\gamma)F_0\\
    F_0
    \end{bmatrix} &= \mathbf{P2VM}\cdot\red{\mathbf{I}} = \mathbf{V2PM}^{-1}\cdot\red{\mathbf{I}} \\
    \gamma &= \mathfrak{R}(\gamma) + i\mathfrak{I}(\gamma)
\end{align}
and finally the squared visibility observable:
\begin{align}
    |V|^2 = \mathfrak{R}(\gamma)^2 + \mathfrak{I}(\gamma)^2
\end{align}

\subsection{Group Delay}

Perhaps one of the biggest advantages in being able to extract the full complex visibility ($\gamma$) for every given frame is \red{that, when spectrally dispersed, we obtain} the ability to extract an estimate of the group delay for each frame. Group delay fringe tracking is the process of measuring the delay offset from a phase of zero through the use of a spectrally dispersed fringe envelope. \red{This can be calculated by taking a Fourier transform of the complex visibilities over the spectral dimension and finding its peak.}

Since we recover the complex visibility, we can take a sum of the complex visibilities over wavelength for a given trial delay, with each $\gamma$ phase-rotated according to the channel's wavelength to compensate for the relative spectral delay. This produces a 'synthetic white-light fringe' for that trial delay\cite{Buscher2015FaintObjects}:
\begin{align}
    \widetilde{F}(x) = \sum_{k=1}^N \gamma_k e^{\frac{i2\pi x}{\lambda_k}}
\end{align}
where the $k$ index is over each spectral channel and $x$ is the given trial delay. 

The group delay is then given by the trial delay where $|F_x|$ is largest. Hence, to find the group delay we must generate a set of trial delays to scan over. We adopted the approach of Basden (2005) \cite{2005BasdenBuscher} in setting the trial delays as:
\begin{align}
\label{eq:trial_delays}
x_p = \frac{p\cdot s}{\Delta \nu} && p=\{-Np+1,-Np+2,... Np\}
\end{align}
where $s$ is a scale factor below unity, $Np$ is the total number of trial delays and $\Delta \nu$ is the wavenumber bandpass of the whole beam combiner:
\begin{align}
    \Delta \nu = \frac{1}{\lambda_\text{min}} - \frac{1}{\lambda_\text{max}}
\end{align}

Thus the group delay estimator can be written as:
\begin{align}
    \tau = \text{argmax}{|\widetilde{F}(x_p)|}
\end{align}
\red{where argmax refers to the function that returns the trial delay for which the flux is maximised.}

\section{Manufacturing}
\label{sect:manufacturing}

\subsection{Method}

We manufactured a triangular tricoupler through the Australian National Fabrication Facility OptoFab node. As our primary use for this beam combiner is the \textit{Pyxis} space interferometer prototype, \red{we designed this tricoupler with the specifications of this interferometer in mind. The science bandpass for \textit{Pyxis}} is 620-760~nm; the lower bound coming from the wavelength cut-off of our 630~nm fibers, and the upper bound at \red{760~nm} from the atmospheric telluric band \red{corresponding to the Fraunhofer A O$_2$ band}. We aimed to manufacture the coupler so that it would provide adequate coupling over these wavelengths. 

As such a device to our knowledge has not been manufactured with these beam combiner constraints, \red{we produced multiple tricoupler devices inscribed with varying parameters to identify which ones resulted in a device with an approximately even splitting ratio over the bandpass}.  The devices were inscribed into Corning Eagle XG boro-aluminosilicate glass via the femtosecond laser direct-write technique \cite{2012flm..book.....O,2015Nanop...4...20G} using a 515~nm laser with a pulse duration of 220~fs at a repetition rate of 1.1~MHz (Light Conversion Pharos). The laser was focused at a depth of 170~\textmu m below the sample’s top surface using a 1.4NA 100x oil immersion objective (Olympus UPlanSApo). At an inscription feed-rate of 250~mm/min and pulse energy of 155~nJ, waveguides were formed that featured optimal mode overlap with a SM600 fiber at 633~nm (4~$\sigma$ waveguide mode-field diameter of 4.5 x 4.1~\textmu m) after the sample was put through a thermal annealing step \cite{2013OExpr..21.2978A} .

\begin{figure}
    \centering
    \includegraphics[width=0.9\linewidth]{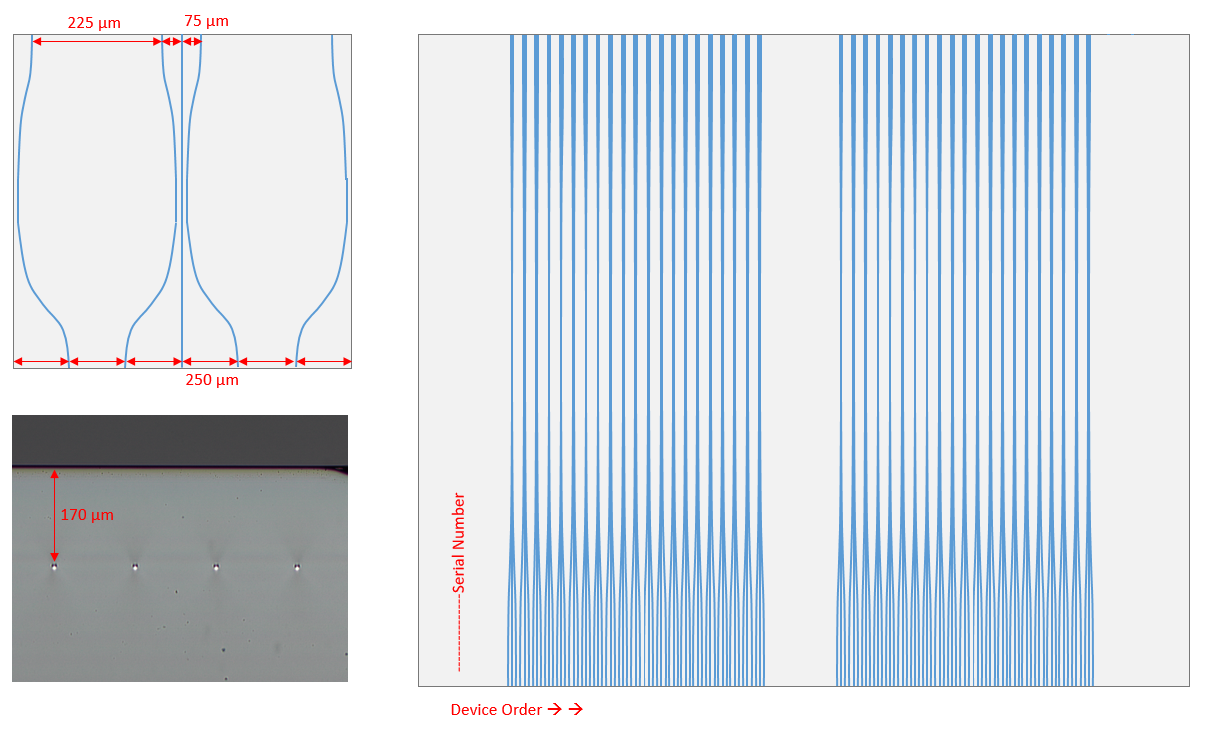}
    \caption{\red{Schematic of the manufactured tricoupler photonic chip.}}
    \label{fig:sketch}
\end{figure}

The tricouplers featured at the output, a pitch of 75~\textmu m between the 3 waveguides arranged in a linear array and at the input, a pitch of 250~\textmu m for coupling to a standard fiber array. \red{A schematic of this design is shown in Figure \ref{fig:sketch}.} Cosine-style S-bends with a minimum bend radius of 20~mm were used to transition the linear arrangement of waveguides at the input and output of the chip to a triangular geometry in the tricoupler’s coupling region. The two outer waveguides formed the base of the nominally equilateral triangle, and the central waveguide was placed at the top vertex of the triangle closest to the top surface with a nominal pitch of 6~ \textmu m between the waveguides. This spacing was selected to ensure the waveguides were inscribed edge-to-edge, \red{to avoid excessive structural overlap that leads to unpredictable coupling and thus, poor reproducibility.}

The tricouplers were optimised in two steps to eliminate asymmetries in the waveguide propagation constants due to the sequential waveguide inscription \cite{2018ApPhL.112k1908D} . First, directional couplers with identical geometry to the tricoupler without their central waveguide were fabricated by changing the feed-rate of the second waveguide in 10 mm/min steps. Asymmetric propagation constants in a directional coupler leads to dephasing that limits the maximum cross coupling to $<$ 100\%. For the particular waveguide pitch of 6~\textmu m, minimum dephasing was observed at a feed-rate of 250 mm/min for the second waveguide, which is identical to the first waveguide’s inscription feedrate. Minimum dephasing enables 100\% power transfer between the waveguides. In the second optimisation step, the third waveguide at the top of the triangle was introduced. Three parameters--the feed-rate of the third waveguide, its vertical position and the length of the coupling region--were varied to obtain close to 33\%/33\%/33\% splitting between the arms. Changing the inscription feed-rate removes the dephasing, i.e. equalises the propagation constant between all 3 waveguides. The vertical position compensates for directional dependent coupling between the waveguides due to the elliptical mode-field profile \cite{2013NatCo...4.1606S} . 

Lastly, changing the length of the coupling region tunes the splitting ratio.  \red{To provide an estimate of the ideal length of the coupling region, we used a photonic simulator, \textit{RSoft}, to calculate the effective indices of the two modes and through equation \ref{eq:theoretical_len}, found an estimated length of 350~\textmu m. This simulation required an approximation for the refractive index profile of the waveguides; we were unable to know this precisely for this photonic chip, and hence used the simulation as an estimate for a length parameter scan.}

\subsection{Characterisation}
\label{sect:characterisation}
To characterise the photonic chip and identify the best performing device from the parameter scan, we devised a phase variance metric derived from the coupling matrix. We linearised the monochromatic intensity as a function of phase by setting $\phi = \phi_0 + \Delta\phi$, giving us:
\begin{align}
    \mathbf{I}(\phi) = \mathbf{I}(\phi_0) + \Delta\phi\dv{\mathbf{I}}{\phi}\left(\phi_0\right).
\end{align}

Assuming we are readout noise limited, which is likely the case in a photon-starved beam combiner, our performance metric, the phase difference variance, can be described by
\begin{align}
    \text{Var}(\Delta\phi) \propto \frac{1}{\sum^3_{j=1}\left[\dv{I_{j}}{\phi}(\phi_0\right)\right]^2},
\end{align}
where the summation index is over the three outputs. The total performance can then be defined by simply summing the variance for each wavelength in quadrature, with the better performer having a smaller variance. This is equivalent to maximising the Fischer information with respect to phase in the readout noise limited regime.

\begin{figure}
    \centering
    \includegraphics[width=0.8\linewidth]{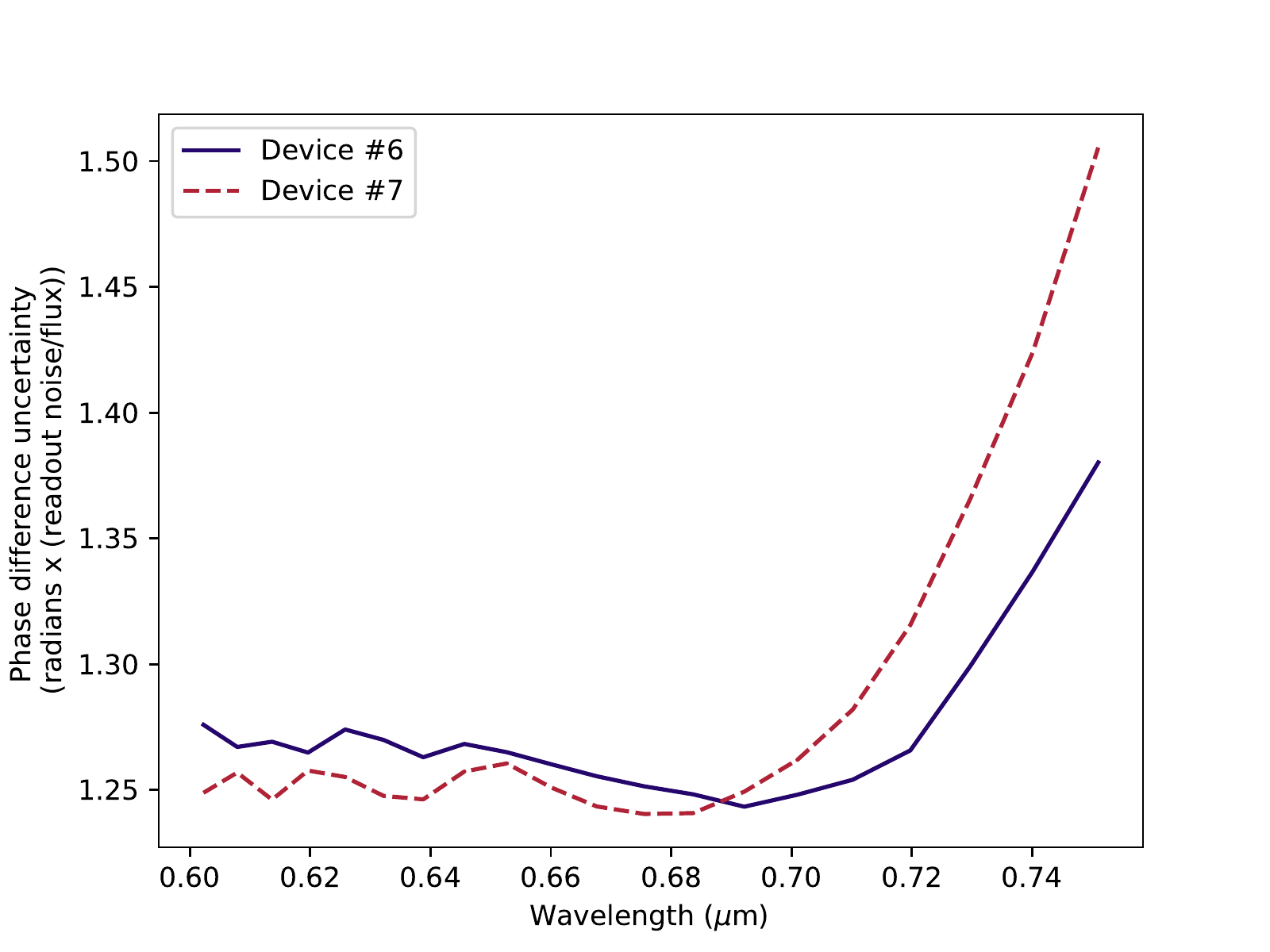}
    \caption{Phase difference uncertainty in the readout noise limit against wavelength for two different tricoupler devices. Note that a smaller uncertainty indicates better performance.}
    \label{fig:phase_variance}
\end{figure}

\begin{figure}
    \centering
    \includegraphics[width=0.8\linewidth]{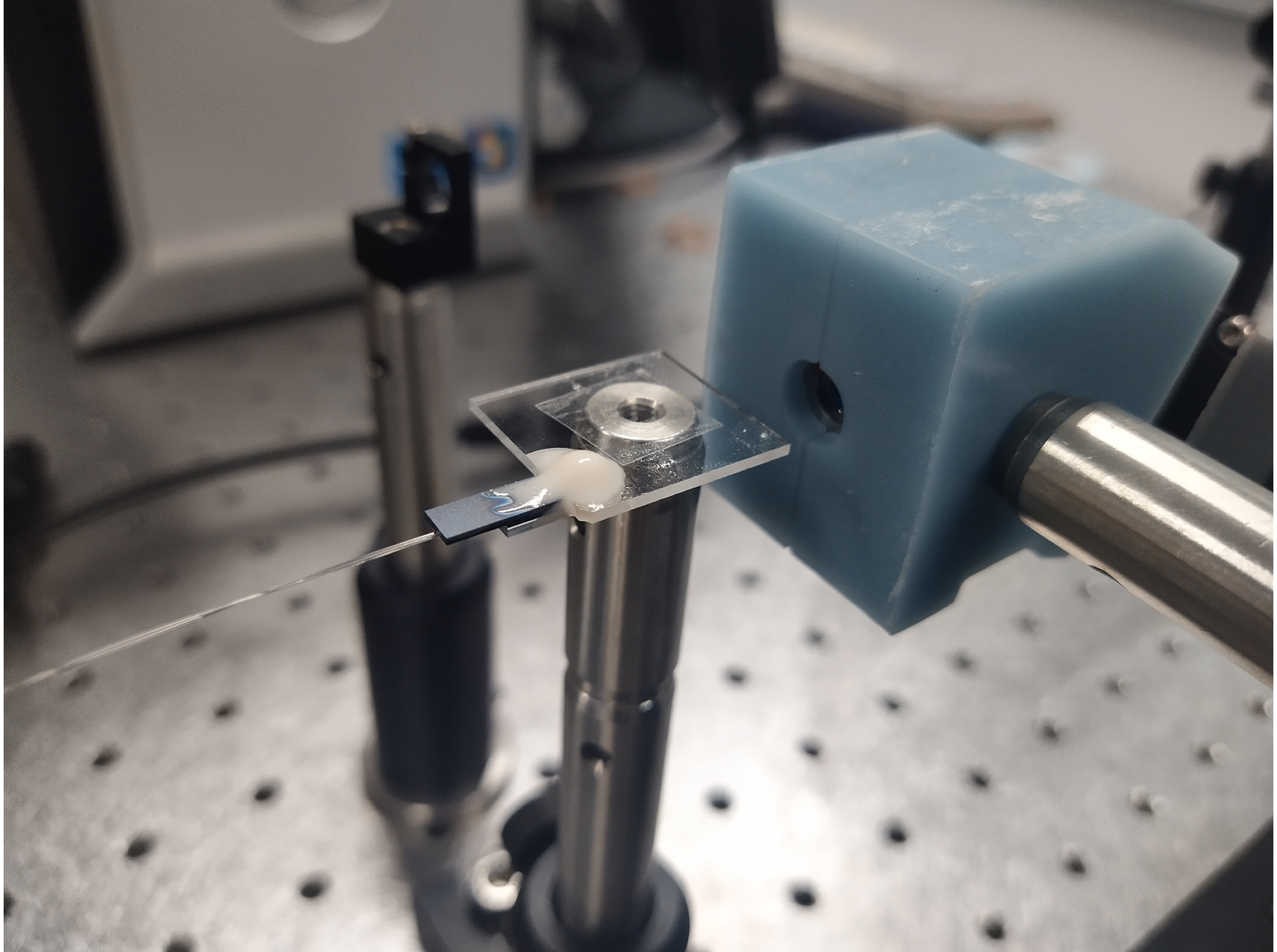}
    \caption{Photograph of the tricoupler photonic chip mounted to a V-groove. The plastic mount holds a low resolution spectrograph ($R\approx80$) and polarimeter for analysis (See Figure \ref{fig:setup}).}
    \label{fig:photo}
\end{figure}

The phase difference uncertainty (square root of the variance) against wavelength for two of the most promising couplers, with approximately even splitting ratios at 660nm, is shown in Figure \ref{fig:phase_variance}. The difference between these two devices is their coupling region: 300~\textmu m compared to 350~\textmu m. Here, we can see that device \#7 performs better at shorter wavelengths, but is worse beyond 700~nm. When summing over wavelength, we in fact see that device \#6 is the best performing over a 620-760~nm bandpass. For this reason, we chose to use device \#6 for our final coupler and as such, carefully mounted and glued a 2-fiber V-groove to the inputs. \red{We used 630-HP polarisation maintaining fiber \cite{630HP} as the inputs to the photonic chip}. A photograph of the chip with mounted V-groove can be seen in Figure \ref{fig:photo}. As we are considering a two telescope beam combiner, the V-groove only allows injection into inputs 1 and 3. For the parameters of the final device, the central waveguide was offset away from coupling region’s central axis by 1~\textmu m in the vertical direction closer to the top surface. An inscription feed-rate of 190~mm/min was used, and the length of the coupling region was 300~\textmu m.

\section{Results}
\label{sect:results}

\subsection{Photometric Performance}
\label{sect:performance}

\begin{figure}
  \centering
    \includegraphics[width=\linewidth]{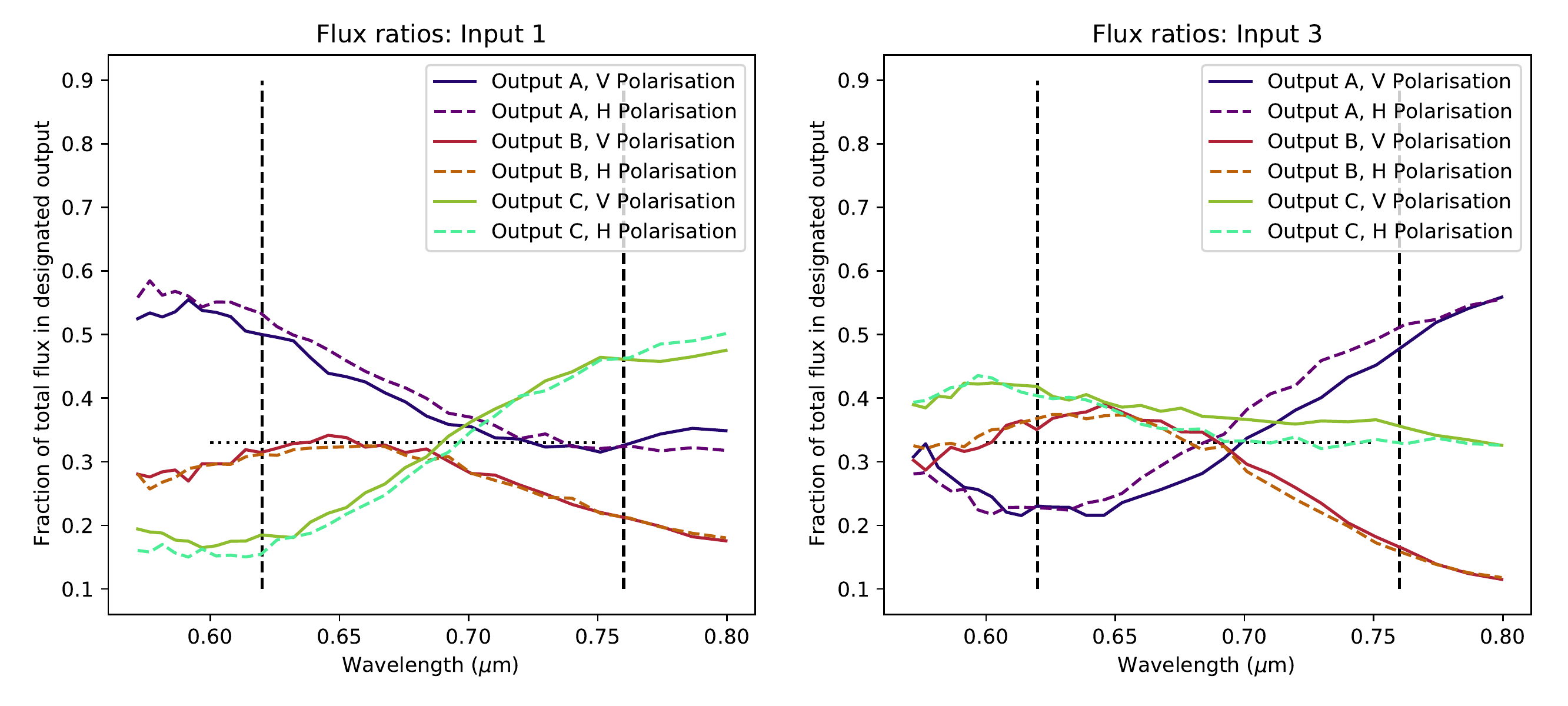}
  \caption{Fractional flux across the three outputs for inputs 1 and 3 of the final tricoupler (device \#6 as explained in Section \ref{sect:characterisation}). Note a Wollaston prism was used to split polarisations; these are denoted as \red{V (vertical) and H (horizontal) respectively.}}
  \label{fig:fluxes}
\end{figure}

A plot of the fractional flux of each output when a broadband light source is injected into a single input for this final coupler is shown in Figure \ref{fig:fluxes}. We did not use input 2 in our performance tests as the V-groove only injects into inputs 1 and 3. Note a Wollaston prism was used to separate polarisations, and are denoted \red{V and H, for vertical and horizontal polarisation} respectively. \red{The polarisation splitter will be used in the final implementation of the \textit{Pyxis} beam combiner in order to calibrate visibilities from polarisation mismatch, as well as providing spectro-polarimetric information on the astrophysical source. For this analysis, however, polarisation effects were not included; as such, when not explicitly mentioned, the vertical polarisation outputs were used}. 

We can see that the coupler performs best around 0.68\textmu m, with performance degrading at higher and lower wavelengths. Nevertheless, the coupling within our defined bandpass remains at worst 51\%:31\%:17\%; leading to a maximum visibility reduction of 13\%. Hence the coupler can adequately function as an interferometric beam combiner.

\begin{table}[]
\centering
\caption{\red{List of measurements for verifying the throughput of the tricoupler device. The different inputs refer to the two different fiber inputs to the tricoupler chip. Measurement units are in $10^{6}$~ADU}}
\label{tab:throughput}
\begin{tabular}{@{}ccc@{}}
\toprule
 & Raw Fiber & Tricoupler \\ \midrule
\multicolumn{1}{l|}{\multirow{4}{*}{Input 1}} & 355 & 319 \\
\multicolumn{1}{l|}{} & 346 & 264 \\
\multicolumn{1}{l|}{} & 346 & 307 \\
\multicolumn{1}{l|}{} & 353 & 309 \\ \midrule
\multicolumn{1}{l|}{\multirow{4}{*}{Input 3}} & 347 & 288 \\
\multicolumn{1}{l|}{} & 359 & 288 \\
\multicolumn{1}{l|}{} & 313 & 263 \\
\multicolumn{1}{l|}{} & 347 & 320 \\ \midrule
\multicolumn{1}{l|}{Mean} & 346 & 295 \\
\multicolumn{1}{l|}{Standard Deviation} & 13 & 21 \\ \bottomrule
\end{tabular}
\end{table}

We also characterised the transmission loss by injecting light from a 660~nm LED through the tricoupler onto a camera, and comparing that to the flux obtained when imaging the same LED through a raw \red{630HP fiber. Four measurements of the output flux for each of the tricoupler inputs were taken alternately with measurements of the raw fiber, using a FLIR Blackfly 05S2M camera to image the flux with a 5~ms integration time. This alternation was used to reduce the effects of changing the fiber connection to the LED source, as well as misalignment issues that may result in changes in flux. Each measurement consisted of taking 100 images and calculating the mean total sum of the set of images after background subtraction. The measured flux values, in units of $10^6$~ADU, are shown in Table \ref{tab:throughput}, along with the mean and standard deviation over the eight measurements. The mean throughput ratio was therefore found to be 85$\pm7$\%. The variance is attributed to the aforementioned alignement and fiber connection variations -} future characterisation will occur when this beam combiner is integrated into the \textit{Pyxis} interferometer and as such we anticipate a more precise measurement at that time. Nevertheless, the throughput is very promising for use in a photon-starved environment. 

\subsection{Interferometric Performance}
\label{sec:int_perf}
\begin{figure}
    \centering
    \includegraphics[width=\linewidth]{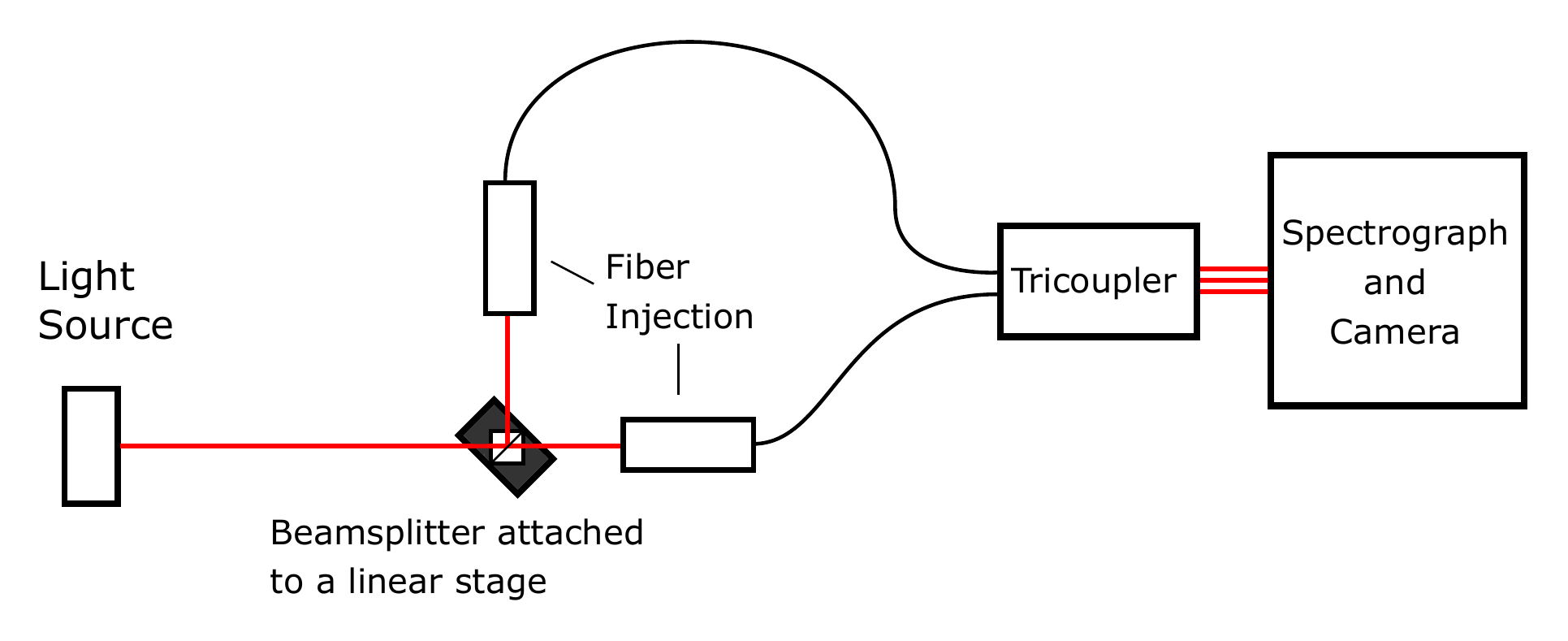}
    \caption{Experimental setup for verifying the interferometric capabilities of the tricoupler.}
    \label{fig:setup}
\end{figure}

Finally, we investigated the response of the coupler with two inputs of differing phase lengths. Broadband light was split through a beamsplitter, mounted diagonally on a linear stage, into the two tricoupler input fibers. In this manner, moving the linear stage changes the optical path difference and hence produce fringes on the output. The out-coming light from the tricoupler was spectrally dispersed and binned into nine channels spanning from 0.61 to 0.76~\textmu m. A schematic of the experimental setup is in Figure \ref{fig:setup}.

\begin{figure}
  \centering
    \includegraphics[width=\linewidth]{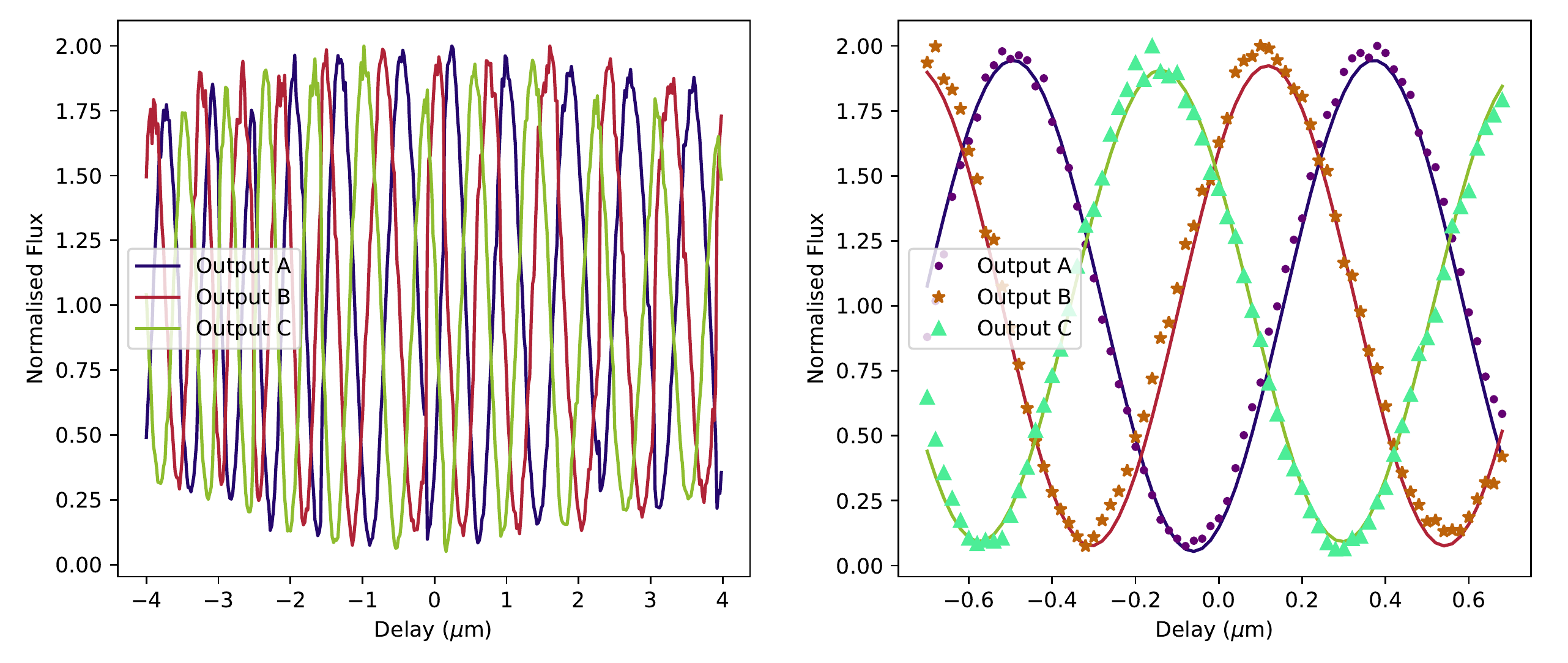}
\caption{Fringe intensity against delay for each of the three outputs of the tricoupler. On the right, in a magnified section of the plot on the left, the intensities at different delays have been fitted with sine curves to emphasise the sinusoidal fringe pattern, high visibility amplitude and the $\frac{2\pi}{3}$ phase shift.}
   \label{fig:fringes}
\end{figure}

A plot of the fringe intensity of each of the three outputs against optical path delay for the channel centred at 0.74~\textmu m is shown in the left panel of Figure \ref{fig:fringes}, where the $\frac{2\pi}{3}$ phase shift between the outputs is immediately apparent. We have also fitted sine curves to the data, shown in the right panel of Figure \ref{fig:fringes}. We observe a good fit to the data, \red{with a mean residual standard deviation of 0.05}, along with a high visibility amplitude of \red{0.93 $\pm$ 0.02} and again the $\frac{2\pi}{3}$ phase shift. 

\begin{figure}
    \centering
    \includegraphics[width=\linewidth]{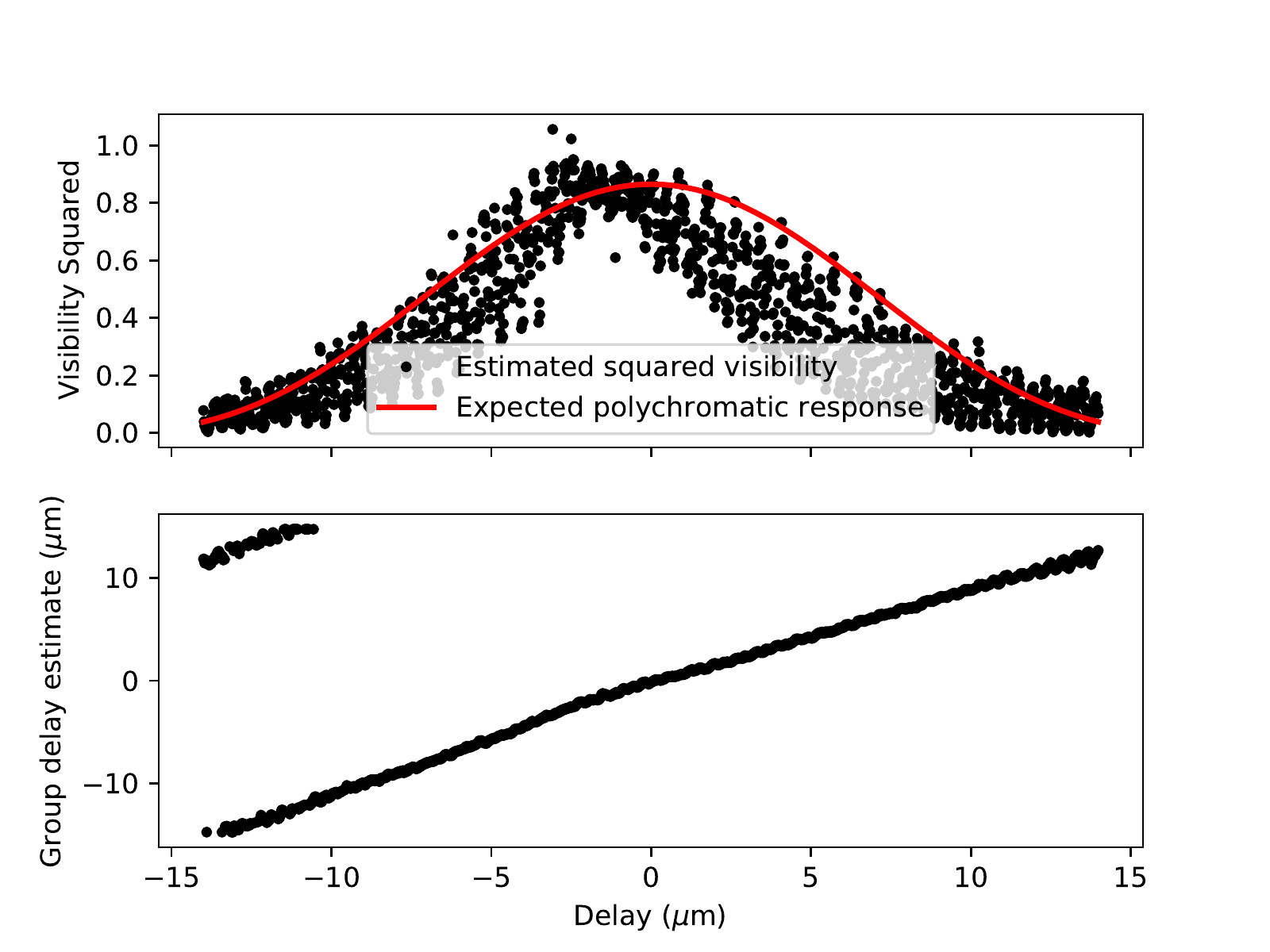}
    \caption{Top panel: Squared visibility response to a change in optical path. Recorded for a wavelength channel centred at 0.74~\textmu m with a width of \red{30\,nm}. The expected response is a squared sinc function as described in Section \ref{sec:int_perf}, scaled by a maximum squared visibility of 0.86. Bottom panel: Group delay estimate for the same changes in optical path.}
    \label{fig:visibilities}
\end{figure}

We then converted the fringe fluxes into visibilities using the P2VM matrix derived in Section \ref{sect:theory}. A plot of the squared visibility for this same wavelength channel as a function of optical path difference is shown in the top panel of Figure \ref{fig:visibilities}. Assuming a rectangular bandpass with no dispersion, the expected polychromatic response, shown in red, is simply a squared sinc function given as:
\begin{align}
    V^2 \red{\propto} \sinc\left(\frac{\Delta \lambda \cdot x}{\lambda^2}\right)^2
\end{align}
with $\Delta \lambda$ being the channel bandpass and $\lambda$ being the central wavelength. \red{Here, the channel bandpass was assumed to be 30~nm; slightly larger than the theoretical channel width due to small amount of defocus in the system.} \red{The squared sinc function was scaled by the visibility calculated earlier of 0.93 (squared visibility of 0.86).}

While the estimated squared visibilities do largely fall on the expected response curve, there is a significant oscillatory \red{data processing noise} signal present as well. This was found to be \red{a byproduct of our estimator}, caused by an uneven flux injected into each input; the greater the uneven injection, the greater the oscillations. \red{As an example, with a flux ratio of 0.8, the amplitude of the squared visibility oscillations is 0.07, whereas a flux ratio of 0.5 results in an amplitude of 0.17.}  We are able to remove this signal through post-processing in a method analogous to Monnier \cite{2001Monnier}; we can use the asymmetries in the coupling matrix averaged over flux phase to recover the intensities, and hence remove the noise contributions. This post-processing procedure will be covered in a future publication. However, note that the fluctuations in $V^2$ in a ground-based interferometer would in practice be smoothed over during natural group delay tracking variations, and still enable robust calibrated visibilities at the 1\% level, \red{competitive with existing visible combiners such as \textit{PAVO}}, by dividing target $V^2$ by calibrator $V^2$ and taking into account a small correction for differences in the overall mean square group delay fluctuation. \red{We also note that the response curve does not perfectly fit the data, particularly around a delay of $\pm 5$~\textmu m; this is due to the spectral bandpass from the detector's pixel response not being perfectly rectangular.} 

\red{We also} recovered an estimation of the group delay for each recorded delay, shown in the lower panel of Figure \ref{fig:visibilities}. \red{We focus on group delay rather than phase delay as, given scintillation and exposure time limitations for \textit{Pyxis}, we do not believe we will be able to track phase delay in real time. For the space-based concept, we will likely need to examine phase delay tracking, but this will be covered in a future publication}. The group delay estimates were recovered using trial delays as described in Equation \ref{eq:trial_delays}, with $N_p = 10^{4}$ and $s = 5\times 10^{-4}$. The delay axis is the same for both panels, showing the correspondence between the squared visibility and estimated group delay. There is also left-right asymmetry present in both the visibility estimate and the group delay - this can likely be attributed to \red{longitudinal} dispersion in the system \red{due to uneven fiber lengths, and could be mitigated through a longitudinal dispersion compensator \cite{1990ApOpt..29..516T}}. We note here that the group delay is only recoverable between a span of $\approx\pm 13 $\textmu m, corresponding to the average coherence length of the wavelength channels:
\begin{align}
    \Lambda = \frac{\lambda^2}{\Delta\lambda} \approx 26 \text{\textmu m}
\end{align}

While the raw estimated visibilities have a large oscillatory signal, the group delay is less affected by this noise source. This highlights one of the main benefits of the tricoupler: we can recover group delay without modulation in real time, while recovering the true visibilities utilising a post-processing algorithm. 

\begin{figure}
    \centering
    \includegraphics[width=0.9\linewidth]{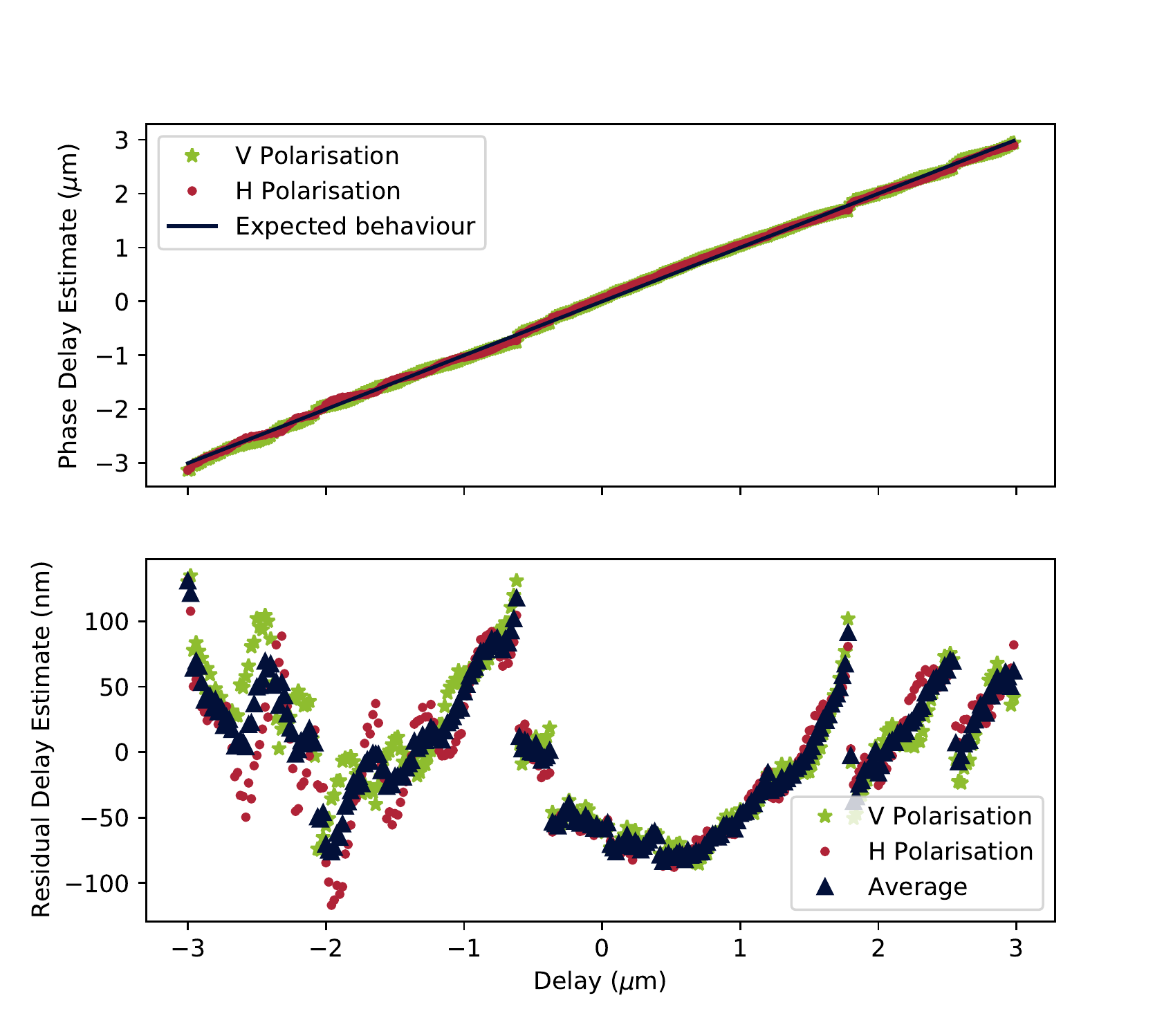}
    \caption{\red{Top panel: Phase delay estimate for changes in optical path, reported for both polarisations. Averaged from all wavelength channels and then scaled for 740nm. Bottom panel: Residuals of the phase delay estimate, in nanometers. The standard deviation of the average of the two polarisations was calculated to be 49~nm.}}
    \label{fig:phase_delay}
\end{figure}

\red{For completeness, we also produced an estimation of the phase delay. This was calculated by taking the average of complex coherences over wavelength, and then taking the phase of this quantity:}

\begin{equation}
    \red{\tau_\text{phase} = \text{arg}\left(\frac{1}{N_\lambda}\sum_\lambda{\gamma(\lambda)}\right)}
\end{equation}

\red{The phases were then unwrapped and plotted against the known linear delay for both polarisations, shown in the top panel of Figure \ref{fig:phase_delay}. To convert between phase and linear delay, we centred the phase at 740~nm, which was the flux-weighted mean wavelength over our wavelength bins. Here, we can see that at least over the inner $\pm$3~\textmu m, the phase delay is quite linear. We also report that each polarisation required a different vertical offsets in order to align with the zero point; this shows us that the combiner produces an inherent instrumental phase shift between polarisations. We calculated this to be 4.74 radians.}

\red{We show the residuals of a straight one to one linear fit in the bottom panel of Figure \ref{fig:phase_delay}, and find that the RMS phase delay error of the average of the two polarisations over this region is 49~nm. For the future space based version of \textit{Pyxis}, in order for phase delay to be useful, we require less than 0.5 radians of RMS error. At 740~nm, this equates to an RMS error requirement of 59~nm, which this chip achieves. We point out though, due a combination of atmospheric scintillation and exposure time limitations, we do not believe we will be able to track phase delay for the \textit{Pyxis} ground based interferometer, especially as it is working at short wavelengths. }

\section{Conclusion}
\label{sect:conclusion}

The triangular tricoupler we have developed has proven to be effective in use as a beam combiner, owing to it having a high throughput of $89\pm11$\% and through maximising visibility information while minimising the number of pixels required. The \red{chip's} ability to obtain the complex coherence instantaneously without path modulation is \red{another} big advantage, particularly for a photon-starved interferometer that needs to maximise integration times. This manifests itself in being able to recover group delay for each recorded frame of delay. We note however that the measured visibilities require a post-processing algorithm to remove a noisy oscillatory signal caused by uneven flux injection.

\red{We note that this work has only been done in the lab, and so the next stage is to perform on sky measurements}. We aim to integrate this chip further into the \textit{Pyxis} interferometer \red{where we will perform on sky tests of the beam combiner in conjunction with the rest of the system. We hope to quantify the performance of the combiner with respect to atmospheric seeing parameters, as well as limiting stellar magnitudes, and ultimately take scientific measurements of objects such as Mira variables.} 

The small size footprint, lack of required modulation for the full complex coherence and simple design \red{of the tricoupler} make it a particularly favourable choice for space interferometry beam combiners, which may hopefully start being developed in earnest in the coming years. 

\subsection* {Acknowledgments}
This research was supported by the ANU Futures scheme and by the Australian Government through the Australian Research Council's Discovery Projects funding scheme (project DP200102383). This work was also performed in-part at the OptoFab node of the Australian National Fabrication Facility, utilising NCRIS and (NSW) state government funding.

\subsection* {Code, Data, and Materials Availability} 

Data and reduction code is available upon request to the author. 


\bibliography{report}   
\bibliographystyle{spiejour}   


\vspace{2ex}\noindent\textbf{Jonah Hansen} is a PhD candidate at the Australian National University, researching into space interferometry. In particular, he is assisting to design and build the Pyxis interferometer - a ground based pathfinder for an eventual space interferometer mission, LIFE (Large Interferometer For Exoplanets), designed to find and characterise exoplanets. He is also assisting in modelling different array architectures and beam combiners for this latter mission. 

\vspace{2ex}\noindent\textbf{Michael Ireland} is a Professor of Astrophysics and Instrumentation Science at the Australian National University. He obtained his PhD from the University of Sydney in 2006, and has since held positions at the California Insititute of Technology, the Australian Astronomical Observatory, Macquarie University and the University of Sydney. His research focuses on stellar astrophysics, exoplanet formation, the search for life on other worlds and technologies needed to support these endeavours

\vspace{2ex}\noindent\textbf{Andrew Ross-Adams} received his Bachelor of Telecommunications Engineering and his Master of Research in Physics from Macquarie University, where he is currently a PhD candidate specialising in femtosecond laser written mode-selective devices, novel techniques for edge coupling to silicon photonics, and configurable integrated spatial division multiplexing for optical fiber communication. 

\vspace{2ex}\noindent\textbf{Simon Gross} received the B.Sc. degree in electrical engineering and M.Sc. degree in microelectronics from Vienna University of Technology, Austria, in 2008 and 2009, respectively. In 2013, he received the PhD in physics from Macquarie University, Australia. He currently holds an ARC Future Fellowship at the School of Engineering at Macquarie University. Simon’s research interests are laser direct-writing of photonic waveguide devices for the use in telecommunication, astronomy, sensing, and information processing applications.

\vspace{2ex}\noindent\textbf{Tony Travouillon} is an associate professor at the Australian National University and is responsible for the instrumentation and telescope development of its school of astronomy. He received his PhD in 2005 from the University of New South Wales.

\vspace{2ex}\noindent\textbf{Joice Mathew} is working as an instrumentation scientist at the Australian National University (ANU). His research interests include electro-optical payload development, instrument modeling, systems engineering, space instrumentation, and qualification. Joice obtained his Ph.D. in astronomical space instrumentation from the Indian Institute of Astrophysics, Bangalore. Before joining ANU, he worked as a visiting instrument scientist at the Max Planck Institute for Solar System Research, Germany on the Solar Orbiter mission.

\vspace{2ex}\noindent Biographies for the other authors are not available.


\end{spacing}
\end{document}